\documentstyle[12pt,epsfig,amsmath]{article}
\newcommand{\m}{\medbreak}

\newcommand{\no}{\noindent}
\newcommand{\EQ}{\begin{equation}}
\newcommand{\eq}{\end{equation}}
\newcommand{\EQA}{\begin{eqnarray}}
\newcommand{\eqa}{\end{eqnarray}}

\newcommand{\AR}{\renewcommand {\arraystretch}{1.5}
\begin{array}{l}}
\newcommand{\bAR}{\renewcommand {\arraystretch}{2}
\begin{array}{l}}
\newcommand{\ARc}{\renewcommand {\arraystretch}{1.5}
\begin{array}{c}}
\newcommand{\bARc}{\renewcommand {\arraystretch}{2}
\begin{array}{c}}
\newcommand{\ar}{\end{array} \renewcommand {\arraystretch}{1}}

\def\ki2{$\chi^2$}
\def\ie{i.e. }
\def\eg{e.g. }

\def\pr#1#2#3{ Phys. Rev. {\bf{#1}} (#2) #3}
\def\prl#1#2#3{ Phys. Rev. Lett. {\bf{#1}} (#2) #3}
\def\pl#1#2#3{ Phys. Lett. {\bf{#1}} (#2) #3 }

\def\ijmp#1#2#3{ Int. J. Mod. Phys. {\bf{#1}} (#2) #3}

\def\apj#1#2#3{ Astrophys. J. {\bf{#1}} (#2) #3}

\begin{document}
\begin{titlepage}
\vspace{0.2in}
\vspace*{-1.5cm}
\begin{center}
{\large \bf On the determination of curvature and dynamical Dark Energy}\\

\m
\vspace*{0.8cm}

J.-M. Virey$^1$, D. Talon-Esmieu$^2$, A. Ealet$^2$, P. Taxil$^1$ and A. Tilquin$^2$

\vspace*{0.5cm}
$^1$Centre de Physique Th\'eorique$^*$,
Case 907,
F-13288 Marseille Cedex 9, France\\
and 
Universit\'e de Provence, Marseille, France\\
\vspace*{0.5cm}

$^2$Centre de Physique des Particules de Marseille,
Case 902,\\
F-13288 Marseille Cedex 9, France
\\

\vspace*{1.cm}

{\bf Abstract} \\
\end{center}

Constraining simultaneously the Dark Energy(DE)
equation of state and the curvature of the Universe is difficult due to strong degeneracies. 
To circumvent this problem when analyzing
data it is usual to assume flatness to constrain DE, or conversely, to assume that DE is a cosmological constant
to constrain curvature.  
In this paper, we quantify  the impact of such assumptions in view of future large surveys. 
We simulate future data for type Ia Supernovae (SNIa), Cosmic Microwave Background (CMB) 
and Baryon Acoustic Oscillations (BAO) for a large range of fiducial cosmologies allowing a small spatial curvature.
We take into account a possible time evolution of DE through a parameterized 
equation of state : $w(a) = w_0 + (1-a) w_a$. 
We then fit the simulated data
with a wrong assumption on the curvature or on the DE parameters.\\
For a fiducial $\Lambda$CDM cosmology, if flatness is incorrectly assumed in the fit and
if the true curvature is within the ranges
$0.01<\Omega_k<0.03$  and $-0.07<\Omega_k<-0.01$, 
one will conclude erroneously to the presence of an evolving DE, even with high statistics.\\
On the other hand, models with curvature and dynamical DE can be confused with 
a flat $\Lambda$CDM model when the fit ignores a possible DE evolution.
We find that, in the future, with high statistics, 
such risks of confusion should be limited, but they are still possible, and biases on the cosmological parameters 
might be important. \\
We conclude on recalling that, in the future, it will be mandatory to perform some complete 
multi-probes analyses, 
leaving the DE parameters as well as the curvature as free parameters. 

\vfill
\begin{flushleft}

Number of figures : 5\\

February 2008, CPT-P008-2008\\

E-mail : virey@cpt.univ-mrs.fr\\

$^*$ ``Centre de Physique Th\'eorique'' is UMR 6207 - ``Unit\'e Mixte
de Recherche'' of CNRS and of the Universities ``de Provence'',
``de la M\'editerran\'ee'' and ``du Sud Toulon-Var''- Laboratory
affiliated to FRUMAM (FR 2291).
\end{flushleft}
\end{titlepage}

   
\pagestyle{myheadings}


\section{Introduction}

Since the discovery of the acceleration of the Universe \cite{Riess98, Perlmutter99} 
many theoretical interpretations
have been developed to find an explanation. These models introduced in general a new 
component called Dark Energy (DE) with an unknown nature \cite{Revues, Copeland06}.  
The distinction between the various interpretations is critical for future cosmology and for
its connections with fundamental physics.

Many studies concentrate on the equation of state (EoS) $w$, defined as the pressure to 
density ratio of this new component.
This parameter is equal to $-1$ for a classical cosmological constant but can be different and/or can 
vary with cosmic time
depending on the various DE models (for a review see \eg \cite{Copeland06}). 
On the experimental side, the determination of the nature of the Dark Energy by constraining 
$w(z)$ is largely debated and should use a large variety of probes and methods \cite{DETF}. 

The WMAP data combined with the recent SDSS data \cite{Eisenstein05, Tegmark07, WMAP3, WMAP5} 
are currently in very good agreement with an adiabatic $\Lambda$CDM model. 
However, most of  studies in the literature  assume a spatial flatness to constrain
the dynamical Dark Energy parameters or, conversely, assume a cosmological constant
to derive constraints on curvature. At the current level of statistical precision, this can be justified 
but various authors have emphasized the dangers one can encounter 
in these too simple strategies \cite{Wright06,Linder06b,Clarkson} 
and the impact of possible biases, when the statistical errors start to be small, 
should be investigated.

In general,  extracting simultaneously $w(z)$ and  the curvature (and also the matter density) 
is not possible since these parameters are perfectly degenerated.  
This is often referred to the so-called ``geometrical degeneracy'' \cite{Bond97,ZSS97,EB99,Huey99}. 
As almost 
all inflationary scenarios predict spatial flatness ($\Omega_k\ < 10^{-10})$, 
this property is often taken for granted in the literature. 
However the precision is not so compelling. 
Experimentally, the error on the curvature parameter $\Omega_k$, as given by the WMAP CMB data, is 
$\pm 0.006 (68\%) ; \pm 0.013 (95\%) $ when adding a Supernovae sample, the Baryon Acoustic Oscillations (BAO) from SDSS
{\em and} assuming a cosmological constant \cite{WMAP5}. 
Assuming 
a parameterized $w(a) = w_0 + (1-a) w_a$ equation of state  (see below),
the precision is only of $\pm 0.018 (68\%) ; \pm 0.04 (95\%) $ \cite{WMAP5,Wang07}. 
Relaxing on the combination with SDSS and the Supernovae's (SNIa), leads even to a precision of $ \pm 0.10 $ \cite{WMAP5}. 
Then, the precision on cosmic curvature is very dependent on the assumption concerning Dark Energy, 
and is strongly limited today by the number of possible free parameters. 

On the other hand, concerning Dark Energy evaluation, even the more recent standard analysis 
as the WMAP ones \cite{WMAP3, WMAP5} or SNLS \cite{SNLS}, 
give results on the Dark Energy parameters using the flatness hypothesis 
and/or no Dark Energy evolution. These analysis use, as is now common, the compressed parameters :
$A$ from BAO \cite{Eisenstein05} and/or of the shift parameter $R$
to include the WMAP CMB data \cite{WMAP3, WMAP5, Wang07, UNION08}. 

Few recent studies \cite{Zhao07,Ichikawa07,Wright07} have extracted simultaneously a free curvature and 
a dynamical Dark Energy  from the present data.  
Such combined analyses 
need always some external priors
or/and assumptions (prior on $h$, the present value of the Hubble constant in unit $100\,
km.s^{-1}.Mpc^{-1}$, assuming massless neutrinos, adiabatic perturbations only...). 
It is known for some time \cite{Maor1} that using incorrect priors  can lead
to an artificial convergence toward a particular model, e.g. the $\Lambda$CDM model, 
(for the consequence of using a strong prior on $\Omega_M$, see e.g. \cite{Virey04b,Upadhye} and \cite{Sahni} for a recent contribution).
The situation could be worse when the number of relevant parameters is increased e.g. when
relaxing a constraint on the curvature and/or the nature of Dark Energy.
\m
In this paper we adopt the same kind of attitude as in \cite{Virey04b, Virey04a}, focusing
on the question of the interplay between the curvature and some DE parameters.
We extend the work of previous authors (e.g. \cite{Linder06b,Clarkson}), 
by addressing these questions more quantitatively and taking into account more realistic observational conditions.

To  do parameter bias estimation with free curvature, which means 
to compare the fitted parameters values to fiducial ones, 
a combined analysis is mandatory. 
We choose to stay in a frame of three combined probes : type Ia  Supernovae (SNIa), 
Cosmic Microwave Background data (CMB) and Baryon Acoustic Oscillations data (BAO).
This combination 
(which is independent of $H_0$) 
has enough statistical power to start to relax on the flatness hypothesis. 
Furthermore, we take advantage on the compressed parameters used in the literature 
as they contain most of the cosmological information needed to reduce Dark Energy parameters errors 
at the same level as from a complete analysis \cite{Wang07}. 
In a simulation study, where all other parameters are fixed, 
this will not introduce additional biases with the main advantage of avoiding 
a complete CMB and BAO data analysis. 
Therefore, we use the $R$ shift parameter \cite{Bond97,EB99,Wang06,Wang07} for the CMB, and the 
so-called $A$ parameter for the BAO \cite{Eisenstein05}. 
This will reduce drastically the computing time, which is useful since
we want to simulate a large range of models, and it will not affect the generality of our analysis.  

In section 2, we present the simulated sample of SNIa, the $R$ and $A$ parameters, 
and the statistical method used for minimization. 
We also introduce a detectability criterion, based on the 
goodness of the $\chi^2$ value of the fit. 

We study in section 3 and 4 two cases which illustrate the
limit of some existing analyses. First, we assume flatness to constrain a dynamical Dark Energy 
in a case of an Universe with a non flat curvature
and a cosmological constant. 
Second, we assume that Dark Energy is a cosmological constant to constrain curvature in
a model of dynamical Dark Energy. 
This case is even more complicated since the number of fiducial parameters 
has increased compared to the previous scenario, this is the "triple trouble" 
situation mentioned in \cite{Linder06b}.

We conclude in section 5 that to get any strong conclusion on Dark Energy evolution 
one will need to fit simultaneously curvature and dynamical Dark Energy parameters
in a multi-probe analyses. 

\section {Cosmological probes, fitting method}

\subsection{The Dark Energy parameterization}

To simulate and fit the involving DE models, we need a parameterization of the EoS $w(z)$. 
We should choose for our study a parameterization that takes into account a possible evolution 
on all redshift range up to the CMB, with a minimum number of free parameters to avoid too large errors.
Indeed it has been shown that at most a two-parameter model can optimally be constrained
by future data \cite{LinderHuterer, Upadhye}.
We choose the Chevallier-Polarski-Linder (CPL) parameterization \cite{Polarski, LinderPRL}, which has been 
extensively used in the literature, in particular by the
Dark Energy Task Force (DETF) \cite{DETF}, as a phenomenological benchmark to compare and contrast the performances
of different DE probes (see e.g. \cite{ealetJMV}):
\EQ
\label{CPLparam}
w(z)=w_0+w_az/(1+z)=w_0+(1-a)w_a
\eq 
$a$ being the scale factor.
The free parameters $w_0$ and $w_a$ characterize the DE model. 
The cosmological constant corresponds to $w_0=-1$ and $w_a=0$. 

Despite its simplicity, this parameterization,
which is bounded for high redshifts 
($w(z \rightarrow \infty )=w_0+w_a$),  fits well the evolution predicted by
a large class of quintessence models (see \cite{CaldwellLinder}
and also \cite{SebJMV} for a recent discussion). 
Note that the CMB analysis introduces in general
a high redshift constraint for the chosen Dark Energy parameterization \cite{Upadhye}, namely
$w(z_{CMB})<0$ at last scattering. 
With our parameterization we should
impose $w_0+w_a<0$ (we refer in the following to this limit as the "CMB boundary condition").
Note however that some DE models can be characterized 
by an EoS such that $w(z_{CMB})\ge 0$. 
This is the case of the early Dark Energy models characterized by a non-negligeable DE density 
at last scattering \cite{Wetterich04} and also for
the tracker quintessence models \cite{Tracker} which have null and positive EoS at early
time as shown explicitly in \cite{CopelandCorasaniti04}. Therefore we will allow to consider sometimes
the part of the parameter space where the "CMB boundary condition" is violated.

Concerning our simulations, this particular choice of the parameterization has no strong impact on the result 
as soon as fiducial models and fitted parameters are using the same modelisation.  
The use of another two-parameter parameterization will not change the main message of this work, we have checked that
it can only slightly modify the size of the errors.

\subsection{SNIa samples}

In the standard Friedmann-Roberston-Walker metric, the apparent magnitude of 
astrophysical objects can be expressed as a function of the luminosity distance:

\EQ\label{mag}
m(z) \; =\; 5\,log_{10}(D_L) + M_B  - 5 \,log_{10}(H_0/c) + 25 \; =\; M_s + 5 \,log_{10}(D_L)
\eq
\noindent
where $M_B$ is the absolute magnitude of SNIa,
$M_s$ may be considered as a normalization parameter
 and 
$D_L(z)\equiv (H_0/c)\;d_L(z)$ is the \emph{$H_0$-independent} luminosity distance to an object 
at redshift $z$. It is given by:
\EQ
D_L(z)=\frac{1+z}{\sqrt{|\Omega_k |}}{\cal S}\left(\sqrt{|\Omega_k |}\,
 \int_0^{z}\,\frac{1 }{E(z')}dz' \right)
\eq
where ${\cal S}(x)=\sinh(x),\; x,\; \sin(x)$ for $\Omega_k>0,\; =0,\; <0$ respectively, with
\EQ
\Omega_k=1-\Omega_M-\Omega_X 
\eq
\no $\Omega_M$ and $\Omega_X$ being respectively the reduced matter and DE densities. We have:
\EQ
  E(z)^2 \; =\;
 \left(\frac{H(z)}{H_0}\right) ^2 \; =\; (1+z)^3\,\Omega_{M}+
    {\rho_X(z)\over \rho_X(0)} \, \Omega_{X}+(1+z)^2\,\Omega_{k},
\eq
\no where 
\EQ
\label{rho}
{\rho_X(z)\over \rho_X(0)} \; =\; \exp \left[ 3\int_0^z\,\left(1+w(z')\right)\, d\,\ln (1+z') \right] 
\eq

Note that we have neglected
the radiation component $\Omega_R $ and this has no impact on our results.
With the parameterization given in eq.(\ref{CPLparam}) the last equation becomes:
\EQ
{\rho_X(z)\over \rho_X(0)} \; =\;(1+z)^{3(1+ w_0 + w_a)} e^{-3w_a z/(1+z)}.
\eq

The studies presented in this paper are performed with simulated Supernovae 
samples with statistics equivalent to what we expect to have
in the future. We concentrate on two sets of data which simulate the statistical
power of the forthcoming and future data.

\begin{itemize}
\item We simulate forthcoming data from ground survey as the large SNLS survey at CFHT \cite{CFHT, SNLS}.
This survey has started in 2003 and the estimation after 5 years of running is to register 
a sample of 700 identified SNIa in the redshift range $0.3<z<1$. 
We simulate a sample as described in \cite{Virey04a}. 
This corresponds to our ``short term'' scenario.
\item We simulate data from a future space mission like JDEM/SNAP \cite{SNAP}, which plans to
discover around 2000 identified SNIa, at redshift  0.2$<z<$1.7 
with very precise photometry and spectroscopy. 
The Supernovae distribution is given in \cite{Kim}.  This corresponds to our ``long term'' scenario.
\end{itemize}
The magnitude dispersion is assumed to be constant and independent
of the redshift at the level of 0.15 for all Supernovae after correction, for both samples. 
We have also neglected systematical errors in this study.

A set of very well calibrated SNIa at redshift $< 0.1$ should be measured by the 
SN factory collaboration \cite{SNFactory}.  This sample is needed
to normalize the Hubble diagram and will be called in the following the "Nearby" sample. 
A "SNAP" ("SNLS") sample means in this paper a simulation of the statistics expected from
the JDEM/SNAP like mission ("SNLS" survey) combined with the 300 (200) nearby SNIa
expected to be measured at the time of SNAP (SNLS).\\

\subsection{The CMB and BAO constraints}

We need to add simulated CMB data to take advantage of the current and future constraints on curvature. 
We have simulated the CMB constraints  by using the CMB shift parameter $R$ \cite{Bond97,EB99,Wang06,Wang07} 
which is the scaled distance to recombination: 
\EQ
R=\frac{\sqrt{\Omega_m}}{\sqrt{|\Omega_k |}}{\cal S}\left(\sqrt{|\Omega_k |}\,
 \int_0^{z_{CMB}}\,\frac{1 }{E(z')}dz' \right)
\eq
where $z_{CMB} = 1089$ is the redshift at the epoch of recombination.
This parameter
contains the full geometrical degeneracy inherent to the CMB \cite{EB99}.
It is measured to be  $R_{WMAP}=1.71 \pm 0.019$ 
from the 5-years WMAP data \cite{WMAP5}.

Using  $R$ we lose some pieces of information which are potentially present in the CMB
but it contains 
enough statistical power to constrain the Dark Energy sector at a level comparable to the one of a full CMB analysis. 
To evaluate the relevance of our results and of this assumption, 
we have made some full CMB calculations and compared the results when evaluating the Dark Energy parameters 
in some models. 
We found a correct agreement between the two methods.
For a comparison between the use of the CMB $R$ parameter and full CMB calculations, one can refer to
\cite{Wang06,Wang07}.

\m
We consider BAO measurements to add an additional constraint on the matter density $\Omega_M$.
We use the so-called $A$ parameter, described in \cite{Eisenstein05}, which encodes almost all information
on the cosmological parameters from present BAO observations :
\EQ
A=\frac{\sqrt{\Omega_m}}{E(z_{BAO})^{1/3}}
\left[\frac{1}{z_{BAO}\sqrt{|\Omega_k |}}{\cal S}\left(\sqrt{|\Omega_k |}\,
 \int_0^{z_{BAO}}\,\frac{1 }{E(z')}dz' \right)\right]^{2/3}
\eq
where $z_{BAO} = 0.35$. It is measured as $A_{SDSS} = 0.469(n_s/0.98)^{-0.35}\pm 0.017\,$ with
$n_s = 0.95$
\cite{Eisenstein05}.
The definition of $A$ is an approximation which incorporates a mixture of the transverse and 
radial information present in the data. \\

In practice we calculate the values of $R$ and $A$ for each simulated fiducial model.
These values are used in the fits along with the errors expected for each observational
scenario.
For the mid term scenario we take the present errors on $R$ and $A$: 
$\sigma (R)=0.019$ (WMAP)\cite{WMAP5} and  $\sigma (A)=0.017$ (SDSS) \cite{Eisenstein05}.
For the long term scenario we choose  $\sigma (R)=0.01$ as roughly expected \cite{Linder06a}
from the Planck mission \cite{Planck}, and  $\sigma (A)=0.005$ as expected for future large surveys 
like a "stage-IV" space based mission dedicated to BAO as described by the DETF \cite{DETF,SPACE}. \\


\subsection{The statistical method}

To proceed in practice, we choose a fiducial model that is a
set of fiducial parameters and simulate data (SNIa magnitudes, $R$ and $A$).
In the most general case the fiducial parameters are 
$\Omega_M^F$, $\Omega_X^F$, $w_0^F$ and $w_a^F$. Note that the value of the SNIa normalisation parameter 
$M_s^F$ is taken to -3.6: its precise value is not much relevant for the following. 
For definiteness we will assume $\Omega_M^F = 0.3$ throughout the paper except otherwise specified but 
we have checked that reasonable deviations ($\pm 0.2$) from this value has a very small influence 
on our conclusions.\\

Then, having chosen a fiducial model, we simulate data for this
model and we fit the simulated data with a wrong assumption on the curvature or the DE parameters.
Finally, we compare the fiducial
and fitted values of the various parameters along
with the respective errors, to estimate how much our conclusions could be affected by
the wrong assumption.

\m

To analyse the (simulated) data,  a minimization procedure has been used \cite{kosmoshow}.
A standard Fisher matrix approach allows a fast estimate of  the parameter errors. 
This method is however limited as it does not 
yield the central values of the fitted parameters.
Then we adopt, unless specified, a minimization procedure based on  a least square method.
The least square estimators are determined by the minimum of the 
$ \chi^2 = ({\bf m} - M(z,\Omega, w)^T {\bf V}^{-1} ({\bf m} - M(z,\Omega, w))$, where
${\bf m} = (m_1...m_n)$ is a vector which contains
the simulated magnitudes plus $R$ and $A$. 
The vector $M(z,\Omega, w)$ is the corresponding vector of values for the fitted parameters
and {\bf V} the covariance matrix.
The errors on the cosmological parameters are estimated
at the minimum by using the first order error propagation technique: $ {\bf U} = {\bf J.V.J}^T $ where {\bf U}
is the error matrix on the cosmological parameters and {\bf J} the Jacobian of the transformation. 
A full $n$-parameter fit ($n$-fit) of the simulated data gives
the central values and errors for the $n$ parameters, along with their correlations
which are in general strong. \\

\subsection{Detectability criterion}

We want to identify among models the more problematic ones where the fit fails
to identify a problem.
Indeed, in any data analysis, a wrong assumption  can be early detected  through a simple $\chi^2$ test:
a high $\chi^2$ indicates that  the {\it goodness of the fit} is bad.
In our context, when fitting simulated data with some hypotheses, a high $\chi^2$
will indicate that some of these hypotheses are wrong.

However, rejecting models from a $\chi^2$ test with real data
is always tricky, therefore we demand a high level of detectability.
Consequently, models where the $\chi^2$ test gives a $5\sigma$ effect,
($\chi^2 >5\sigma (\chi^2 )$) where the
{\it rms} of the $\chi^2$ is $\sigma(\chi^2)=\sqrt{2N_{dof}}$ and
where $N_{dof}$ is the number of degrees of freedom in the fit,
are considered to be easily identified as in conflict with the input hypothese.

On the other hand some models  passing this test can be biased as the fitted parameters are too far from
the fiducial ones.
In the next sections, we identify the models which yield a good value of the $\chi^2$ 
even if the fitting hypotheses are incorrect. A careful attention is given to the resulting
confusions and biases.

\section{Reconstruction of DE dynamics with a wrong assumption on curvature}

If we consider realistic experimental conditions, the reconstruction of an evolving Dark Energy is difficult
when we leave the curvature free due to a complete degeneracy between the relevant parameters. 
This comes from the integral relation between cosmological distances and the cosmological parameters 
\cite{Maor1} and the only way to solve this question is to fix some parameters 
or/and to combine different informations. 
However, such a strategy can lead to some biased interpretations. 
This  has recently been studied on theoretical grounds in \cite{Clarkson}. 
Following the same approach, we simulate fiducial models as  $\Lambda$CDM Universes with non-zero curvature and 
we calculate the errors on the cosmological parameters as expected in future observational 
conditions. 
In this way we are able to quantify when some models with curvature, consistent with the CMB constraint, 
can start to bias the results on the DE EoS when flatness is incorrectly assumed in the fit.
In practice, we fix $\Omega_M^F$ in the range $0.1-0.5$ 
(illustrations are given for $\Omega_M^F=0.3$)
and we vary $\Omega_X^F$ to scan $\Omega_k^F$ between -0.1 and 0.1.
Then we perform the fit of $\Omega_M, w_0$ and $w_a$, assuming flatness.

To test the wrong hypothesis we first test the {\it goodness of the fit} 
as described in the previous section.
When fiducial curvature is too far from flatness, the $\chi^2$ is too high
and the wrong assumption is detectable. 
Models with $\Omega_k$ lower than -0.08 (-0.07) and $\Omega_k$ greater than 0.05 (0.03)
for the short term (long term) scenario are detected by the simple $\chi^2$ test.
We study in more details some models which give good $\chi^2$ fits to determine the risks 
of a wrong conclusion on the values of the DE parameters. 
\begin{figure}[h]
\centerline{\includegraphics[height=10.truecm,width=15.truecm]{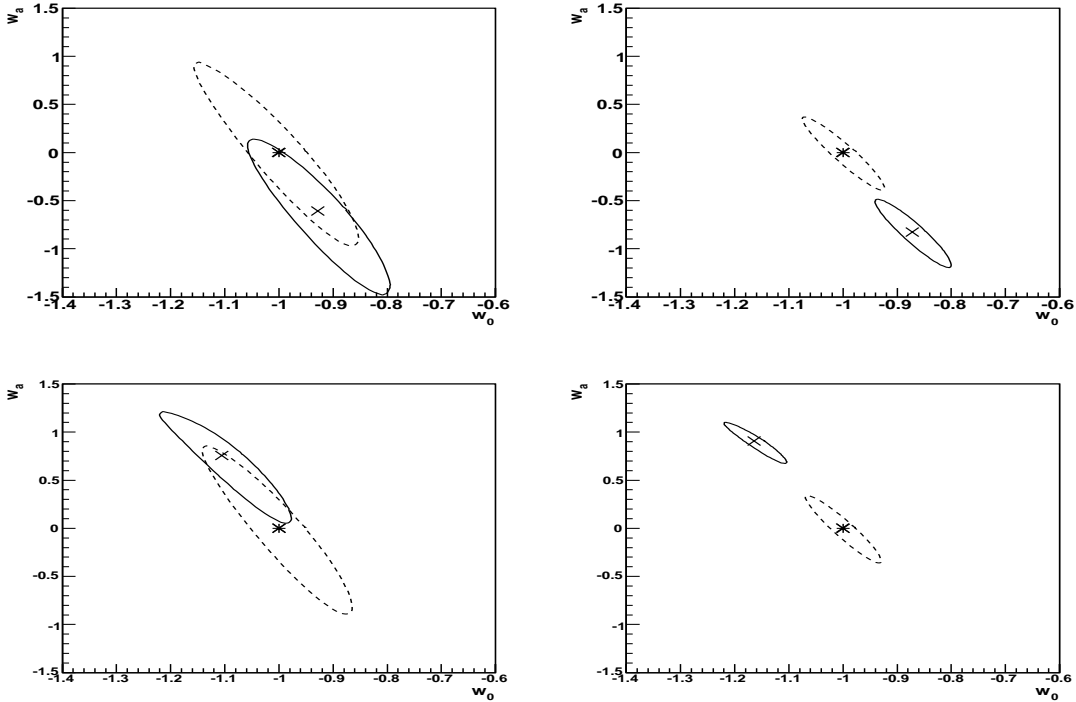}}
\caption{Illustration of the bias problem in the $(w_0,w_a)$ plane
for a slightly open model (upper plots) with $\Omega_M^F = 0.3, \Omega_X^F = 0.68$ ($\Omega_k^F = 0.98$) 
and for a slightly closed model (lower plots) with $\Omega_M^F = 0.3, \Omega_X^F = 0.72$ ($\Omega_k^F = 1.02$). Results
are given for a short term scenario (left plots) and a long term scenario (right plots).
A fiducial $\Lambda$CDM model has been assumed. Contours are at 1$\sigma$ and correspond
to some fits with (solid) or without (dashed) the flatness assumption.}
\label{fig1}
\end{figure}

In Figure 1 we present 2 examples of fits with good $\chi^2$ in the ($w_0, w_a$) plane. 
We have chosen two different fiducial models: a $\Lambda$CDM open model (upper plots) with
($w_0^F = -1$, $w_a^F = 0$, $\Omega_M^F = 0.3$, $\Omega_X^F = 0.68$) and a 
$\Lambda$CDM closed model (bottom plots) with
($w_0^F = -1, w_a^F = 0, \Omega_M^F = 0.3, \Omega_X^F = 0.72$).
In each case the full simulated data (SNIa, $R$ and  $A$), 
are fitted with the (wrong) hypothesis of flatness. The $1\sigma$ contours correspond
to the solid curves and the left (right) plots correspond to the short term (long term) scenario.

One can see that the fitted $w_0$ and $w_a$ central values are far from the fiducial values 
in each case. 
We get for positive curvature ($\Omega_k<0$) a $w_0$ value 
which is pushed towards the phantom regime $w_0<-1$
and a $w_a$ value which is positive, and the converse for negative curvature ($\Omega_k>0$)
\ie that $w_0>-1$ and $w_a<0$.
Let us stress again that this bias is not detectable with true data since the
fit is good (low $\chi^2$). 
Note that in the same time the $\Omega_M$ value is well reconstructed.\\

For the long term scenario, the statistical power increases, then the error ellipses
are obviously smaller but the bias is higher.
For comparison we show the contours resulting from 
a complete fit where the curvature is a free parameter (dashed contours). 
These contours are centered on the fiducial values as expected. 
In the long term scenario the contours from the two approaches are clearly disconnected,
hence the problem becomes stronger.
\m

Let us quantify more precisely the biases which are expected on these two DE EoS parameters.
We vary the curvature 
and determine the zones of bias, defined as $\mid w_0 +1\mid > \sigma(w_0)$ and 
$\mid w_a \mid > \sigma(w_a)$. 
The DE parameters bias means that a cosmological constant with some amount of curvature  
will be interpreted as a dynamical DE model in a flat Universe.
The results are summarized in Table 1: 
\begin{itemize}
\item In the positive curvature sector, $-0.08 < \Omega_k^F < -0.02$ (short term),
$-0.07 < \Omega_k^F < -0.01$ (long term). The upper bound is the bias limit, above this value the
results are not biased. The lower bound is the {\it goodness of the fit} limit, below this value the
$\chi^2$ explodes.
\item In the negative curvature sector, $0.02 < \Omega_k^F < 0.05$ (short term),
$0.01 < \Omega_k^F < 0.03$ (long term): now the lower bound is the bias limit and the upper bound
the goodness of the fit limit.
\end{itemize}

\begin{table}
\begin{center}
\begin{tabular}{|c|c|c|c|c|c|}
\hline
 & Bad Fit & Biased Zone & Validity Zone & Biased Zone & Bad fit\\
\hline
short term & $\Omega_k^F<$-0.08& -0.08$<\Omega_k^F<$-0.02 & -0.02$<\Omega_k^F<$0.02 
& 0.02$<\Omega_k^F<$0.05 & $\Omega_k^F>$0.05\\
\hline
long term &$\Omega_k^F<$-0.07 & -0.07$<\Omega_k^F<$-0.01& -0.01$<\Omega_k^F<$0.01 
& 0.01$<\Omega_k^F<$0.03& $\Omega_k^F>$0.03\\
\hline
\end{tabular}
\end{center}
\caption{Validity range of the flatness assumption for the determination of the
DE EoS parameters $w_0$ and $w_a$.}
\end{table}

In other words, if we do not take into account the current and future curvature uncertainties, 
results on the DE parameters can be biased even in a long term scenario. 
This implies a potential misinterpretation of the DE dynamics 
even when  using a combination of precise probes both at low and high redshift, 
as soon as a spatial flatness is assumed.

We conclude that $w(z)$ can be reliably  measured with the flatness assumption only if
the true curvature is $|\Omega_k| \leq 0.02$ $(\leq 0.01)$ for a short (long) term scenario.\\

\section{Reconstruction of curvature with a wrong assumption on DE}

Another common strategy is to pin down the curvature of the Universe 
when assuming that DE is a cosmological constant ($\Lambda$CDM hypothesis).
In this section we will investigate the biases introduced by this assumption. 

After a short illustration of the problem, we study in details the possible confusion
of some non-flat dynamical DE models with a flat cosmological constant.
Then we quantify the models which can provide a biased estimation of the cosmological parameters
in an undetectable way.\\

\subsection{Illustration}

In this study, we test the $\Lambda$CDM model assumption by a fit of $\Omega_M$ and $\Omega_X$.
First we perform the fit for two examples which illustrate the situation.
We consider two models with a dynamical DE for two different curvatures ($\Omega_k^F = 0.03$ and 
$\Omega_k^F = -0.02$), still fixing the matter density to  $\Omega_M^F = 0.3$. 
\\
We show on Figure 2 the results of the fits in the $(\Omega_M, \Omega_X$) plane 
for the combination of SNIa, R et A in the short term scenario. 
The straight line corresponds to flat models ($\Omega_M + \Omega_X = 1$).
We show in solid curves the contours obtained from a fit assuming that
DE is described by a $\Lambda$CDM model. 
We see that the fitted values of the matter and Dark Energy densities (crosses) are far from the
fiducial ones (stars) due to the wrong hypothesis.
Therefore the fitted values of curvature are compatible with the flatness at $1\sigma$.
For comparison the fits without assumption on the DE are performed and the results are plotted with 
the dotted contours. 
Note that these ellipses are completly disconnected. 
These results are given for a short term scenario but this bias still exists for the long term
scenario.\\

\begin{figure}[h]
\centerline{\includegraphics[height=8.truecm,width=16.truecm]{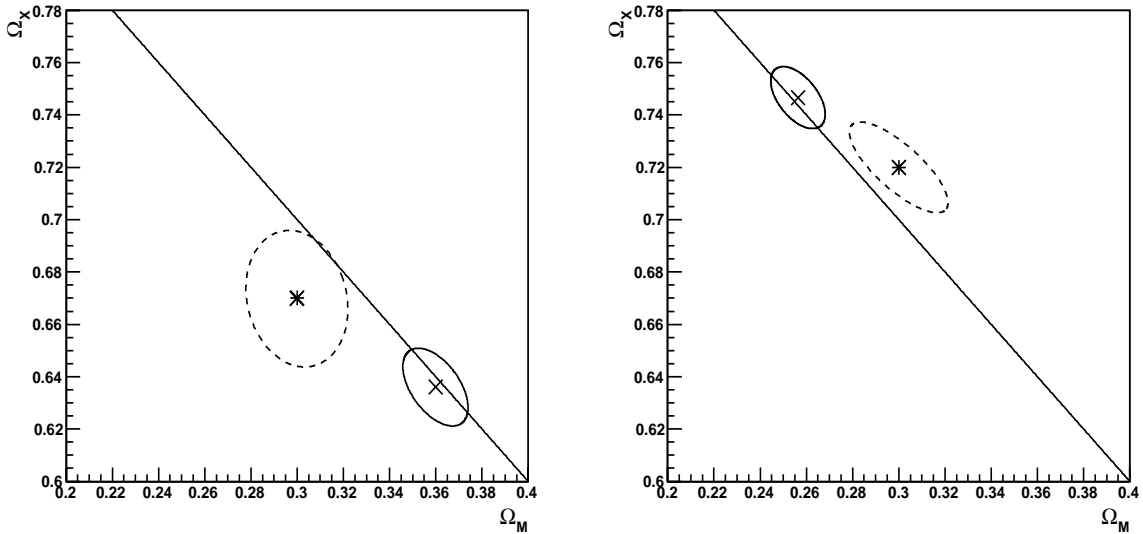}}
\caption{Contours at $1\sigma$ in the ($\Omega_M ,\Omega_X$) plane for a short term scenario. 
The left (right) plot corresponds to the
fiducial cosmology $\Omega_M^F=0.3$, $\Omega_X^F=0.67$, that is $\Omega_k^F = 0.03$, $w_0^F = - 0.9$ 
 and $w_a^F = 0.5$
($\Omega_M^F=0.3$, $\Omega_X^F=0.72$, that is $\Omega_k^F = -0.02$, $w_0^F = - 0.9$ and $w_a^F = -1.5$).
Plain (dotted) curves are obtained from fits with (without) the assumption that DE is
a cosmological constant.
The straight line corresponds to flat universes.}
\label{fig2}
\end{figure}

With this illustration we have shown that it is possible to interpret a dynamical DE model in a Universe
with a non-zero spatial curvature as a flat cosmological constant model, for present 
statistics as well as for
future surveys. 
In the following we look for the DE models and the possible amount of curvature
where such a misinterpretation of combined data is possible.\\

\subsection{Confusion with flatness}

We want to identify the DE models with non zero curvature which can be compatible with 
 a flat $\Lambda CDM$ model, when assuming incorrectly $w=-1$ in the fit.

For a given set of fiducial parameters  $(\Omega_k^F,w_0^F,w_a^F)$ 
we perform a fit imposing $w=-1$.
For each tested model we first verify the {\it goodness of the fit} and then  we determine if the fitted value of the
curvature is compatible with the flatness by satisfying $|\Omega_k|<\sigma (\Omega_k)$. 

\begin{figure}[h]
\vspace*{-0.4truecm}
\centerline{\includegraphics[height=10.truecm,width=10.truecm]{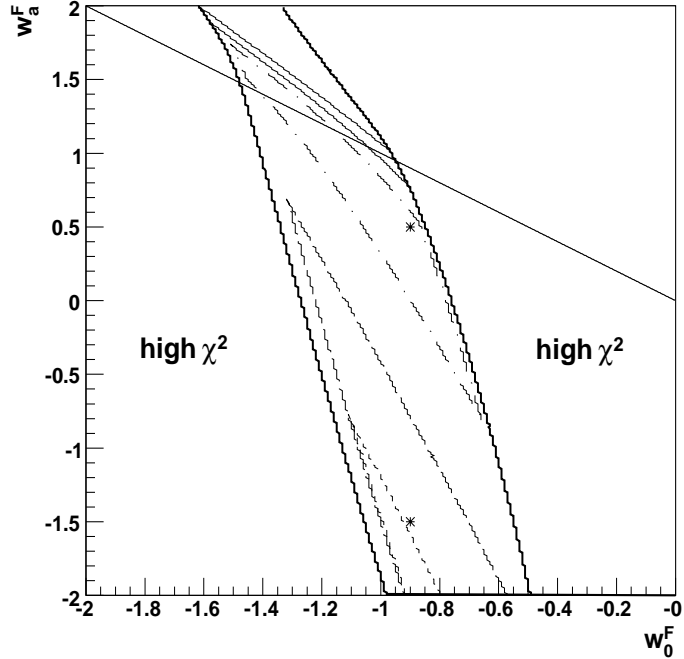}}
\caption{The black contour gives the limit of goodness of the fits.
Other contours correspond to the zones of confusion with a flat $\Lambda$CDM model
for the following fiducial cosmologies ($\Omega_M^F=0.3$): $\Omega_X^F=0.64$
($\Omega_k^F = 0.06$) (plain),
$\Omega_X^F=0.68$ ($\Omega_k^F = 0.02$)(dotted-dashed), 
$\Omega_X^F=0.72$ ($\Omega_k^F = -0.02$)(long-dashed) and
$\Omega_X^F=0.73$ ($\Omega_k^F = -0.03$)(short-dashed). The straight line is the "CMB boundary
condition" $w_0+w_a<0$. Calculations are done for the short term scenario.}
\label{fig3}
\end{figure}

We present in Figure \ref{fig3} the results for four fiducial curvatures ($\Omega_k = 0.06, 0.02, -0.02,
-0.03$) in the short term scenario.
Models with a good fit are located between the two solid curves and represent a large area of the plane.
It means that many DE models in a curved space may give a correct fit for a cosmological
constant.
For some of these models it exists a confusion with a flat Universe, \ie these non-flat models are 
interpreted as flat ones.
For each fiducial curvature these models form a {\it confusion zone} in the plane. 
Note that the models chosen for the illustration of the problem on Figure 2 
are marked by  stars on Figure 3 and are within the {\it confusion zones}.

\begin{figure}[h]
\centerline{\includegraphics[height=10.truecm,width=10.truecm]{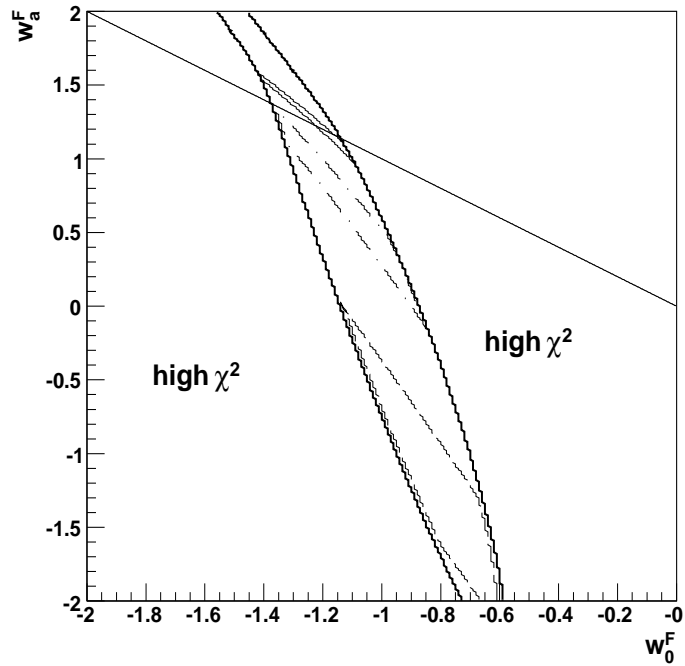}}
\caption{The black contour gives the limit of goodness of the fits. 
Other contours correspond to the zones of confusion with a flat $\Lambda$CDM model
for the following fiducial cosmologies ($\Omega_M^F=0.3$): $\Omega_X^F=0.67$ ($\Omega_k^F = 0.03$)(plain),
$\Omega_X^F=0.69$ ($\Omega_k^F = 0.01$)(dotted-dashed) and
$\Omega_X^F=0.71$ ($\Omega_k^F = -0.01$)(dashed). The straight line is the "CMB boundary
condition" $w_0+w_a<0$. Calculations are done for the long term scenario.}
\label{fig4}
\end{figure}

We have performed the same study for the long term scenario. 
The results for fiducial curvature ($\Omega_k^F$) between -0.01 and 0.03 are given in Figure \ref{fig4}.
Thanks to the higher statistics, the area of the ($w_0^F,w_a^F$) plane where the fit passes the $\chi^2$ test is smaller.
Therefore the wrong assumption is detectable for a greater number of models.

Concerning models which pass the $\chi^2$ test but may be confused with a flat cosmological constant model,
we remark that DE models with positive curvature ($\Omega_k^F<0$) 
are in the lower part of the plane (roughly $w_0^F>-1$ and $w_a^F<0$) on both Figures 3 and 4. 
With high statistics, most of these problematic models exit the {\it zone of
goodness of the fit} and can be rejected as soon as $\Omega_k^F<-0.01$. 

The DE models with negative curvature ($\Omega_k^F>0$) that may be confused with
a flat cosmological constant model are in the upper part of the plane
(roughly $w_0^F<-1$ and $w_a^F>0$).
Note that some {\it confusion zones} are close to, or above, the "CMB boundary condition".  
If we ignore this limit, some very open models 
(up to $\Omega_k^F = 0.6$) can still give a correct $\chi^2$ and
be confused with a flat cosmological constant Universe.
However, for these models, the size of the {\it confusion zone} is very tiny and corresponds to a very small 
phase space volume.

Therefore, our analysis cannot exclude that some open dynamical DE
models with a large departure from flatness can induce a confusion with a flat $\Lambda CDM$ model.
The proximity or violation of the CMB boundary condition indicates that these DE models 
are rather exotic: of both phantom and early Dark Energy types. 
The smallness of the phase space volume for these models show also that these exotic DE models should
be fine tuned to give a confusion.
This situation means that our analysis
has reached its limits and that a better description of the DE model (beyond the CPL parameterization)
and of the CMB (beyond the use of $R$) should be done to yield a confirmation of these
possible confusions. 

One can find in the literature constraints on curvature coming from different combined analyses where
the DE dynamics is left free \cite{Zhao07,Ichikawa07,Wright07}. 
They give $\Omega_k<0.02$ at 95\%CL, concluding that open models are disfavored.
Our last result indicates that one has to be cautious: constraints depend of the statistical methods which are
very sensitive to the size of the phase space volume \cite{Upadhye} and the current combined analyses of data
may have neglected or missed these open exotic DE models.\\

In summary, even with a large statistical sample, it is still possible to confuse a dynamical DE non-flat model 
with the flat $\Lambda$CDM model when a cosmological constant is assumed. 
This is due to the geometrical degeneracies between the DE and the CMB parameters and this cannot be completely solved 
by the use of a combined analysis with SNIa and BAO, even at high statistics. 
To draw any conclusion on the description of Dark Energy, it is mandatory to fit enough DE parameters with the curvature as a free
parameter.

\subsection{Bias on the reconstruction of the reduced densities}
In this section, we study the biases on the reduced densities $\Omega_M$ and $\Omega_X$, still
in the context of fiducial curved dynamical DE models analyzed 
assuming a cosmological constant. 
For that purpose,  we vary $\Omega_M^F$ in the range  0.1 to 0.5
and  $\Omega_X^F$ in a range
such that
$|\Omega_k^F|<0.2$. Then, we scan the ($w_0^F,w_a^F$) plane.
We quantify the biases introduced by the wrong hypothesis,
when trying to reconstruct  the cosmological parameters $\Omega_M$ {\it and} $\Omega_X$.
We define the bias of a parameter $p$ by $B_p=|p^F-p|$ and we say that $p$ is biased (valid) if the bias
is larger (smaller) than the error obtained for $p$ \ie if $B_p>\sigma (p)$
($B_p<\sigma (p)$).
(One can consult \cite{Virey04a} for more details
on these definitions).\\
There are some  zones where the $\chi^2$ is too high, then the fit is not good and the problem is detectable.
When the $\chi^2$ is good
we can define two types of zone :

\no - the  zone where the $\chi^2$ is good and the fitted parameters are well reconstructed even with the assumption
of $w=-1$, we call it the {\it Validity Zone} (VZ).
This zone is centered on the $\Lambda$ point ($w_0^F=-1$, $w_a^F=0$) where the fit assumption is true.

\no - the zone where the $\chi^2$ is good but all fitted parameters are wrong\footnote{For the short term
scenario another situation may appear where $\Omega_k$ is well-reconstructed but $\Omega_M$ and $\Omega_X$
are biased, namely the biases compensate to give accidentally the correct curvature.
This possibility disappears at high statistics.}.
The previously discussed {\it zones of confusion} belong to this zone and correspond to the
cases where we obtain erroneously $\Omega_k=0$. \\

Figure 5 shows one example where $\Omega_M^F=0.25$ and $\Omega_X^F=0.7$ in
the long term scenario.
The result is generic for all models considered in this study and whatever the statistic is.
We see that, with the combined analysis and high statistics, there are large zones in the plane
where the cosmological parameters are wrong, if we assume DE to be a cosmological constant.
This fact is well known.
The new information is that many models will be interpreted as a  $\Lambda CDM$ because
of the degeneracy of curvature and dynamical Dark Energy
and this will not be detectable.

\begin{figure}[h]
\centerline{\includegraphics[height=10.truecm,width=10.truecm]{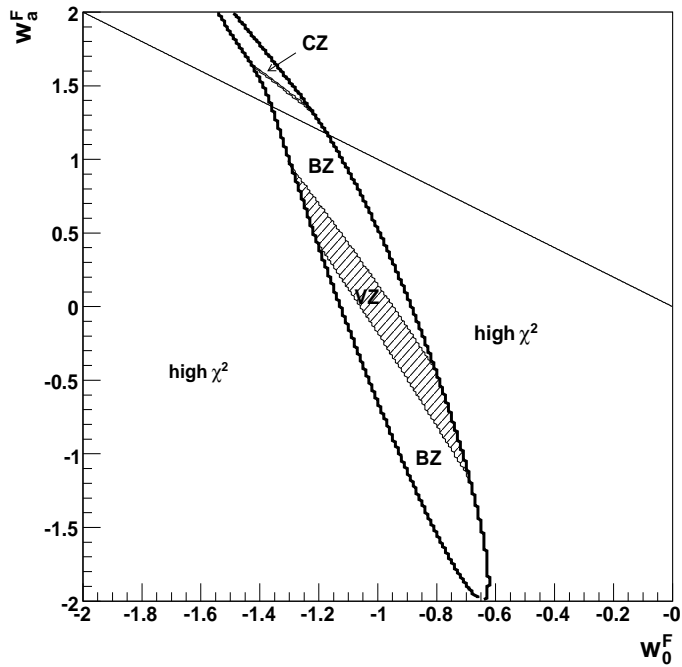}}
\caption{The black contour gives the limit of goodness of the fits.
Other contours correspond to the zones of confusion with a flat $\Lambda$CDM model
(labelled by CZ),
the validity zone (labelled by VZ) and the biased zone (labelled by BZ).
The fiducial cosmology is $\Omega_M^F=0.25$ and $\Omega_X^F=0.7$ ($\Omega_k^F = 0.05$).
The straight line is the "CMB boundary
condition" $w_0+w_a<0$. Calculations are done for the long term scenario.
}
\label{fig5}
\end{figure}

The position of the various zones in the fiducial plane can move when fiducial
parameters like $\Omega_M^F$ or/and $\Omega_X^F$ are changing:\\
\no - An increase of $\Omega_X^F$ reinforces
the power of the $\chi^2$ test (\ie the goodness of fit zone decreases)
but reinforces also the bias problem
(\ie the VZ decreases).\\
- Changing  $\Omega_M^F$  shifts the location of the various zones in the fiducial plane.
\\

Scanning all possibilities,  we can always find some values for the densities where the $\chi^2$ is
low, and the fitted results are valid or biased. 
Our study shows also that the addition of SNIa and BAO
reduces the degeneracies but do not eliminate them.

Finally, let us notice that we find also that assuming only a constant $w$, different from $-1$, leads
to similar confusion and biases.

\section{Conclusions}
The vicious circle of assuming flatness to constrain dark
energy, or conversely, of assuming that dark energy is a cosmological constant 
(or have a constant EoS) to constrain curvature, should be avoided. 
Indeed, such assumptions may provide
very important biases and confusion if they are not true.

We have used a combined analysis using SNIa, BAO and CMB data, taking avantage of the use of 
the compressed parameter $A$ for BAO and shift parameter $R$ for CMB. 
The combination  will help to break degeneracies, origin of the biases, which come mainly  
from  the integral relation between cosmological distances and the cosmological parameters \cite{Maor1}.

We see that assuming flatness, a cosmological constant model 
with a small amount of curvature can be
interpreted as a dynamical DE model in a flat universe. 
We find that, even using a multi-probe approach
and with future high statistics, such hypothesis will bias 
the interpretation of the data if the true curvature is in the range 
$0.01<\Omega_k <0.03\, $ or $\,-0.07<\Omega_k <-0.01$.
This is due to the fact that, even with small statistical errors, this range of non-flat models
will give a good $\chi^2$ when assuming flatness, because of the degeneracy with the
DE sector. We emphasize that these models are always inside the current level of uncertainties of WMAP data, 
(when an evolving DE is assumed in the analysis).
Let us also state that the PLANCK mission will not allow to reduce strongly these uncertainties as the information on the curvature
is mainly coming from the first peak  which is already quite precisely determined. 

Conversely, if we suppose that Dark Energy is a cosmological constant,
a dynamical non flat DE model may be interpreted as a flat cosmological constant model.
We show that for closed models, there is no confusion as soon as  $\Omega_k> -0.01$.
For open models, there is a greater risk of confusion, independently of the statistical sample,
in particular if $0.01<\Omega_k<0.05$. However, the confusion
may still happen at $\Omega_k>0.05$, but only
for very exotic DE models (of both phantom and early
DE types) which seem less favored by the current data.

In addition, we show that in case of misinterpretation, when assuming a cosmological constant,  
the cosmological parameters $\Omega_M$ and $\Omega_X$ will be biased and this will be undetectable in the fit for a large
range of models.
In any case, high statistics (long term scenario) help, in the sense that they allow to detect
more easily a wrong hypothesis through the $\chi^2$ test. However, there is still a number of 
DE models where biases will not be detected and the smaller errors would increase our confidence
in a wrong interpretation.\\

In general, in most of the analysis of real data, two fits are performed independently : one with the assumption  of flatness
and the other with the assumption  of a Cosmological constant.
Consistency between the two fits is then considered as a proof a reliability.
However, this approach is not exempt of drawbacks. 
For example, in the spirit of our analyses of Section 3, if we vary the DE parameters
beyond a simple cosmological constant we find some confusions zones which overlap the ones of Figures 3 and 4. This means that
these models yield a good $\chi^2$ in all cases and point toward a flat $\Lambda$CDM. In these cases the circular logic
\cite{Clarkson, Wright06} of the "consistency" test fails.\\

The strategy which consists of relaxing the curvature and the DE parameters within one global procedure 
will be the only one which avoids confusion. In addition, when comparing 
the errors expected on the cosmological parameters in the fit procedures
with or without the wrong assumptions (see the plain and dashed contours
of Fig. 1 and  Fig.2), one sees
that in a combined analysis using SNIa+CMB+BAO, there is no gain of assuming flatness
or/and that DE is a cosmological constant.
This fact has been already noticed, see \eg \cite{Linder06b}.

In conclusion, our quantitative analysis allows us to strongly support
the points already stressed by a few authors : it is imperative
to let the curvature and the DE parameters as free parameters. This is already true for
the analysis of present data as well as for data expected from the future large surveys.
\\

\no {\bf Acknowledgments}\\ 

Special thanks to C. Marinoni and C. Tao for helpful comments and suggestions.
A.T. and J.-M.V. thanks the CNRS/DRI and the Chinese Academy of Sciences for financial support.
In particular, we thank 
 Xinmin Zhang (IHEP), Hong Li (Peking University), 
Gong-Bo Zhao (IHEP)
and Qin Bo (NAOC) for interesting discussions and their very kind hospitality
at their Institutes in Beijing where the present work has been started.
Finally, we are indebted to Hong Li for performing full CMB calculations to check some of our results.


\begin{thebibliography}{99}

\bibitem{Riess98} A.G. Riess et al., Astron. J. {\bf 116}, 1009 (1998)

\bibitem{Perlmutter99} S. Perlmutter et al. Astrophys. J. {\bf 517}, 565 (1999)

\bibitem{Revues} P.J.E. Peebles and B. Ratra, Rev. Mod. Phys. {\bf 75}, 559 (2003);
T. Padmanabhan, Phys. Rep. {\bf 380}, 235 (2003).

\bibitem{Copeland06} E.J. Copeland, M. Sami and S. Tsujikawa, \ijmp{D15}{2006}{1753}

\bibitem{DETF} Dark Energy Task Force report to the Astronomy and Astrophysics  
Advisory Committee, http://www.nsf.gov/mps/ast/detf.jsp ; A. Albrecht et al., astro-ph/0609591

\bibitem{Eisenstein05} D.J. Eisenstein et al., Astrophys. J. {\bf 633}, 560 (2005)

\bibitem{Tegmark07} M.Tegmark et al. Phys. Rev. D  {\bf 74} 123507 (2006) [astro-ph/0608632]

\bibitem{WMAP3} D.N. Spergel et al. Astrophys.J.Suppl. {\bf 170} 377 (2007) [astro-ph/0603449]

\bibitem{WMAP5} E. Komatsu et al. [arXiv:0803.0547] \\
http://lambda.gsfc.nasa.gov/product/map/current/parameters.cfm

\bibitem{Wright06} E.L. Wright, astro-ph/0603750, Talk given at the Int. Astrophys. Conf. on Relativistic Astrophysics and Cosmology : Enstein's
Legacy, Munich, November 2005

\bibitem{Linder06b} E.V. Linder, Astropart.Phys. {\bf 26}, 102 (2006) 

\bibitem{Clarkson} C. Clarkson, M. Cortes and B.A. Bassett, JCAP {\bf 0708} ,011 (2007) 

\bibitem{Bond97} J.R. Bond, G. Efsthatiou and M. Tegmark, 
Mon. Not. R. Astron Soc. {\bf 291}, L33 (1997)

\bibitem{ZSS97}  M. Zaldarriaga, D.N. Spergel and U. Seljak, Astrophys. J. {\bf 488}, 1 (1997)

\bibitem{EB99}  G. Efsthatiou and J.R. Bond, 
Mon. Not. R. Astron Soc. {\bf 304}, 75 (1999)

\bibitem{Huey99} G. Huey et al., Phys. Rev.{\bf D59}, 063005 (1999)

\bibitem{Wang07} Y. Wang and P. Mukherjee, \pr{D76}{2007}{103533}

\bibitem{SNLS} P.Astier et al,  Astron. Astrophys.  {\bf 447}, 31 (2006)

\bibitem{UNION08} M. Kowalski et al. [arXiv:0804.4142]

\bibitem{Zhao07} G.B. Zhao et al., Phys. Lett. {\bf B648}, 8 (2007)

\bibitem{Ichikawa07} K. Ichikawa and T. Takahashi, JCAP {\bf 0702}, 001 (2007).

\bibitem{Wright07} E.L. Wright, \apj{664}{2007}{633}.

\bibitem{Maor1} I. Maor, R. Brustein and P.J. Steinhardt, Phys. Rev. Lett.
{\bf 86}, 6 (2001); I. Maor et al., Phys. Rev. {\bf D65}, 123003 (2002).

\bibitem{Virey04b} J.-M. Virey et al.,  Phys.Rev. {\bf D70}, 121301 (2004)

\bibitem{Upadhye} A. Upadhye, M. Ishak and P.J. Steinhardt, Phys.Rev. {\bf D72}, 063501 (2005) 

\bibitem{Sahni} V. Sahni, A. Shafieloo and A. A. Starobinsky, 2008 [arXiv:astro-ph/0807.3548] 

\bibitem{Virey04a} J.-M. Virey et al.,  Phys.Rev. {\bf D70}, 043514 (2004)

\bibitem{Wang06} Y. Wang and P. Mukherjee, Astrophys. J. {\bf 650}, 1 (2006)

\bibitem{LinderHuterer} E.V. Linder and D. Huterer, Phys. Rev.{\bf D72}, 043509 (2005).

\bibitem{Polarski} M. Chevallier and D.  Polarski, Int.J.Mod.Phys. {\bf D10}, 213 (2001)

\bibitem{LinderPRL} E.V. Linder, Phys. Rev. Lett. {\bf 90}, 091301 (2003)


\bibitem{ealetJMV} J.M. Virey and A. Ealet, Astron. Astrophys. {\bf 464}, 837 (2007)

\bibitem{CaldwellLinder} R.R. Caldwell and E.V. Linder, Phys. Rev. Lett. {\bf 95}, 141301 (2005)

\bibitem{SebJMV} S. Linden and J.M. Virey, Phys. Rev. {\bf D78}, 023526 (2008).

\bibitem{Wetterich04} C. Wetterich, \pl{B594}{2004}{17}.

\bibitem{Tracker} R.R. Caldwell, R. Dave and P.J. Steinhardt, \prl{80}{1998}{1582}.
    
\bibitem{CopelandCorasaniti04} P.S. Corasaniti and E.J. Copeland, \pr{D67}{2003}{063521}.

\bibitem{CFHT} see e.g. 
http://cfht.hawai.edu/Science/CFHTLS-OLD/history$\_$2001.html

\bibitem{SNAP} http://snap.lbl.gov

\bibitem{Kim} A.G. Kim et al., Mon. Not. R. Astron Soc. {\bf 347}, 909 (2004)

\bibitem{SNFactory} W.M. Wood-Vasey et al., New Astronomical Review {\bf 48}, 637 (2004)




\bibitem{Linder06a} E.V. Linder, Gen. Rel. Grav. {\bf 40}, 329 (2008)

\bibitem{Planck} http://planck.esa.gov

\bibitem{SPACE} A. Cimatti et al. [arXiv:0804.4433]
%
\bibitem{kosmoshow} Our simulation tool, the ``Kosmoshow'', has been
developped by A. Tilquin and is available upon request or directly
at\\
http://marwww.in2p3.fr/renoir/Kosmoshow.html.


\end{thebibliography}
\end{document}